# Visual novelty, curiosity, and intrinsic reward in machine learning and the brain

Andrew Jaegle, Vahid Mehrpour, Nicole Rust
Department of Psychology, University of Pennsylvania

## Abstract

A strong preference for novelty emerges in infancy and is prevalent across the animal kingdom. When incorporated into reinforcement-based machine learning algorithms, visual novelty can act as an intrinsic reward signal that vastly increases the efficiency of exploration and expedites learning, particularly in situations where external rewards are difficult to obtain. Here we review parallels between recent developments in novelty-driven machine learning algorithms and our understanding of how visual novelty is computed and signaled in the primate brain. We propose that in the visual system, novelty representations are not configured with the principal goal of detecting novel objects, but rather with the broader goal of flexibly generalizing novelty information across different states in the service of driving novelty-based learning.

## Highlights

- Novelty-based exploration can expedite learning when rewards are sparse
- Novelty-based machine learning incorporates novelty into computations of value
- In brains and machines, novelty signals are continuous and distributed
- Effective novelty-based machine learning requires view and state invariance
- IT cortex supports flexible view and state invariant representations of novelty

**Funding:** This work was supported by the National Eye Institute of the US National Institutes of Health (grant number R01EY020851) and the Simons Foundation (through an award from the Simons Collaboration on the Global Brain).

**Introduction**

What signals does an animal or other agent need to learn to generate good behavior? Reinforcement learning (RL) provides one answer to this question: to learn to act, an agent needs access to a reward signal, and it needs to estimate which states and actions cause that signal to be large [reviewed by 1]. Typically, RL assumes that the reward signal comes in the form of an external reward, such as a satisfying food. RL has proven highly effective for describing reward-based learning in neural systems [2], and it has led to breakthroughs in machine learning [3,4]. However, classic implementations of RL are most effective when reward signals are easily and frequently obtained [1], and this is typically not the case in real environments. What type of signal can be used to learn in settings where rewards are rare?

Curiosity in the form of novelty seeking is one candidate for such a signal [5]. Human infants exhibit strong preferences for stimuli and situations unlike those encountered before, even when rewards aren't otherwise associated with these stimuli [6]. Analogous novelty preferences have been observed in over 100 species, ranging from reptiles to monkeys [7,8]. Recent work in machine learning demonstrates how novelty can be incorporated into RL algorithms to successfully drive the mastery of complex tasks. In parallel, recent developments in neuroscience have provided important insights into how visual novelty is computed and signaled in the brain. In this review, we synthesize recent work on the computation, representation, and uses of visual novelty in both machine learning and neuroscience. By drawing these connections, we hope to encourage more rapid progress in both fields.

**Visual novelty and intrinsic motivation for machine-based RL**

Consider a monkey foraging for fruit hidden in the dense brush of a tree, where different branches can be envisioned as pathways that lead to different "states" where the fruit might be (Fig 1). How might the monkey decide which states are valuable? When fruit is sparse, the monkey is unlikely to find rewards by exploring branches purely at random, as only a few states contain fruit, and it will typically miss these states, often revisiting fruitless parts of the tree. A more effective strategy is to explore branches that are different than those tried in the past. This can be achieved by making the act of exploring novel things itself a rewarding experience for the monkey, and this is precisely what novelty-driven RL algorithms seek to do.

Classic RL algorithms learn what outcomes lead to what rewards by assigning values to different states. Novelty-based RL methods seek to incorporate the novelty of a state into its value by augmenting the classic equation for computing value, the Bellman equation [9], with a third term that reflects the state's novelty:

$$V(s) = max_{a \in A}[R(s,a) + \gamma \mathbb{E}_p[V(s')] + \beta N(s)]$$

where $V(s)$ is the value of the current state, $s$. The first term of this equation, $R(s,a)$, is an estimate of the reward that the agent will get by taking the most rewarding action ($a$) from the set of possible actions ($A$). The second term, $\mathbb{E}_p[V(s')]$, is the anticipated value of the next state, $s'$, taking into account the expected dynamics of the environment (described by a transition function $p(s'|s,a)$). Here $\gamma$ is a discount factor used to balance the reward from the current state with the expected value of future states. Together, these first two terms capture the original Bellman formulation, which does not incorporate novelty considerations. The final term, $N(s)$, was incorporated later to assign value based on the state's novelty, where $\beta$ is a parameter that balances reward seeking with novelty seeking [10].

In simple environments, novelty can be computed by keeping track of all states that have been visited, and states are rated as more novel if they have been visited less frequently. Algorithms that employ such an approach are often called *count-based strategies.* In early proposals for novelty-driven RL, count-based strategies were demonstrated to lead to good exploration behavior in simple RL settings such as bandit problems and simple Markov decision processes [11-13].

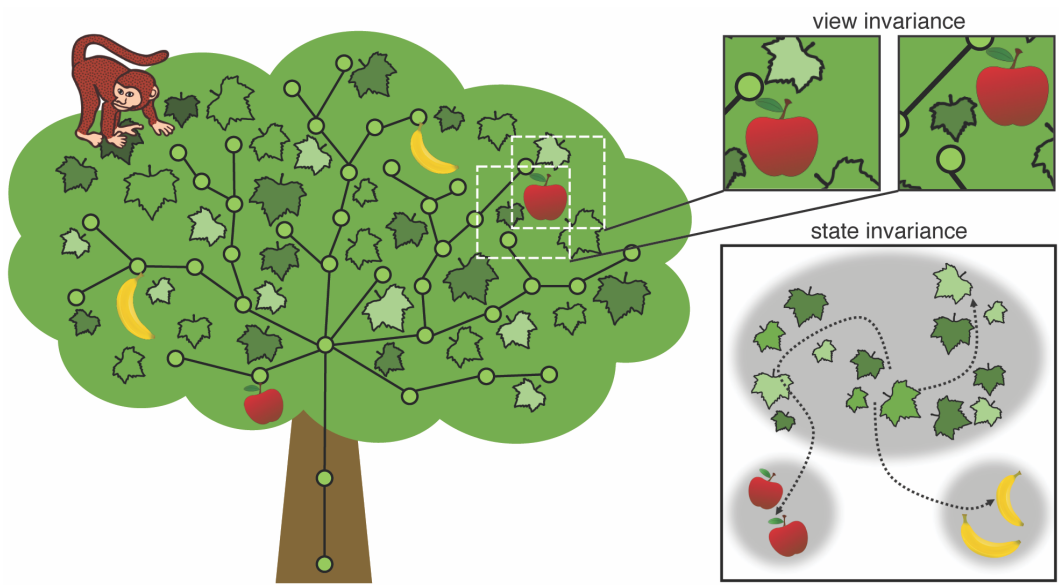

**Figure 1:** *Novelty can drive exploration in environments with sparse external rewards.* An illustration of the benefits of exploration: a monkey is trying to find a piece of fruit (a reward) in a large, densely foliated tree with many branches. Typically, the monkey must make many choices and explore many "states" before it receives a single reward. If the monkey finds novel states rewarding, then it will be encouraged to explore the tree, and it can discover rewarding states that it would otherwise miss. *View invariance:* different views of the same state (e.g. the apple) can correspond to different images. To effectively drive RL, a system must map images onto their corresponding states. *State invariance:* different states can share features that are indicative of their novelty, e.g. reflecting the fact that fruits are usually large and are rarely green while leaves are often small and can take on many shades of green. A system that can exploit the features shared by different states can drive the monkey to explore states with novel features (e.g. objects with a new size or shade).

Contemporary RL approaches typically address the problem of training an agent to produce a controller or motor command given the pixel pattern on a video display, in settings such as Atari video games [14] or navigation through complex 3D scenes (Beattie C *et al.*, arXiv: 1612.03801). To effectively estimate and use novelty in these settings, a system must overcome the high-dimensionality inherent to these problems, a challenge that can be parsed into two conceptually distinct invariance problems. The first is *view invariance*: because the same underlying state can be viewed from many different vantage points, a system has to sort out the mapping from a diversity of images on to their corresponding states to estimate the novelty of the current visual input (Fig 1). However, this alone is not sufficient to explore settings with a large number of states, where exhaustive exploration of unique states remains intractable. Rather, a system must also be capable of *state invariance*: capturing the groupings of different states based on their shared characteristics, such as the fact that ripe fruits are rarely green (Fig 1). In contemporary novelty-driven RL algorithms, state-invariant estimates of novelty are a crucial component of guiding exploration toward states that are most unlike those previously seen and efficiently finding previously undiscovered rewards.

Recent novelty-driven RL methods address both invariance problems by employing some form of a model that transforms images into novelty estimates. Most methods train a deep neural network (DNN) for this goal, as DNNs have proven successful at learning invariant and generalizable representations in a broad range of settings (for reviews, see [15,16]). Recent proposals differ in terms of the details by which these models are trained and how the resulting measures of image similarity are converted into estimates of novelty, as described below. Once estimated, nearly all proposals incorporate novelty into measures of value in the same way (i.e. by incorporating novelty into the Bellman equation), as described above.

One class of recent proposals for novelty-driven RL addressed the view and state invariance challenges by incorporating a method for computing similarity between images. Some of these proposals extend classic count-based proposals by first mapping similar images to the same bin (e.g. using a hash function [17] or by clustering [18]) before counting. A complementary proposal [19] estimated image similarity by training a DNN to estimate the distance in time between pairs of images, motivated by the observation that temporal continuity can capture invariance [20]. Instead of counting how many times states had been visited, this method estimated an image's novelty by comparing it to familiar images held in a memory buffer.

Another class of proposals for novelty-driven RL used "pseudocounts" to estimate novelty. In contrast to explicit count-based proposals, pseudocount methods address the invariance problems using a model (e.g. a DNN) trained to estimate the probability of images. Pseudocounts are computed based on how the probability of an image changes between the Nth and (N+1)th times it is viewed, where N is often zero (see [21] for the exact expression). Because these methods approximate continuous probabilities, they are well suited to scenarios where the dimensionality is high and hence nearly all counts are zero, but some stimuli are more probable than others. Because pseudocount methods are based on the response after repeated exposure to an image, they bare some similarity to the phenomenon of *repetition suppression* in the visual system (see below for a discussion of the role of repetition suppression in novelty computation in the brain). Several recent papers have achieved promising results using pseudocounts to drive exploration on difficult RL tasks [21-23].

How does novelty seeking relate to other types of intrinsic motivation? In the case of pseudocount estimates, an important theoretical link has been established between estimates of novelty and the amount of *information gain* [24] that follows from observing an image [21]. Intuitively, this is because the novelty of an image reflects how much it differs from what we expect to see based on what we've seen in the past. Early work on intrinsic motivation established the link between curiosity and measures of image informativeness, such as information gain [25,26]. While the information gained by observing an image cannot be computed directly, several methods have been proposed to approximate it to drive exploration [27,28]. A number of recent papers have also proposed other, related signals to drive exploration, including methods that estimate image informativeness by how variable or unreliable the response to the image is [29], how well the response of a target model can be predicted [30], and by how difficult it is to predict what will follow an image in time [31-33]. Unlike methods for novelty-driven RL, these methods do not estimate novelty explicitly, but they share the goal of driving exploration by estimating the informativeness of images.

In sum, machine learning algorithms that incorporate novelty as an intrinsic reward have proven effective at driving RL, particularly in environments where rewards are sparse. To compute novelty, contemporary algorithms are designed to generalize across different views of the same state and to share features across different states (Fig 1). These algorithms estimate the novelty of the current visual input using a variety of techniques, but we can broadly distinguish between those that employ explicit count-based methods of novelty and those that estimate novelty using pseudocount procedures. Direct theoretical links have been established between pseudocounts and the amount of information gained by viewing an image. All methods share the goal of driving exploration toward states that have not yet been adequately explored. In the next section, we focus on recent progress toward understanding where and how the neural correlates of these machine-based novelty signals may be computed and signaled in the primate brain.

**Computing and signaling novelty in the primate brain**

Our ability to detect visual novelty (or equivalently remember whether we have seen an image) is quite remarkable – for example, we can view thousands of photographs, each only once and each for a few seconds, and then distinguish with high accuracy the specific images that we have already seen from those that remain novel to us [34]. Even after viewing as many as 10,000 distinct photographs, our rates of remembering do not saturate, suggesting that this type of single-exposure visual memory has an exceedingly large capacity [34,35]. The most remarkable aspects of our ability to detect visual novelty are thought to be mediated by our "familiarity" memory system. One effective illustration of familiarity is the experience that we all occasionally have of seeing someone and remembering that we know them but not being able to recall any details about them, at least for a few moments, and this "sense of remembering absent details" is precisely what the familiarity memory system supports. In contrast to recollection-based memories (e.g. of the details about that person, such as their name), which are thought to be largely mediated by the hippocampus, familiarity is thought to be mediated by another brain area in the medial temporal lobe, perirhinal cortex, as well its input from the part of the visual system involved in signaling object and scene information, inferotemporal cortex (IT) [reviewed by 36,37].

How do these brain areas signal visual novelty? Novelty is thought to be signaled in IT and perirhinal cortex via an adaptation-like change in firing rate in response to familiar as compared to novel stimuli, a phenomenon referred to as *repetition suppression* [38-42]. Consistent with the signatures needed to account for the vast capacity of human single-exposure visual memory behavior, firing rate reductions with familiarity are selective for images, even after viewing large numbers of them, and these response decrements last for several minutes to hours following the single viewing of an image [39,40,42]. These putative visual novelty signals are mixed with signals reflecting visual identity, both within the responses of individual neurons and across the IT population. That is, visual identities of images and their content are thought to be reflected as distinct patterns of spikes across the IT population, and this translates into a population representation in which visual information about the currently-viewed scene is reflected by a population's vector angle [Fig. 2; reviewed by 43]. In contrast, novelty is thought to be signaled by overall firing rates or equivalently the length of the population response vector, where vectors are longer for novel images and become shorter as they become familiar [42]. Novelty and familiarity modulations are thought to be approximately multiplicative [44,45], which translates to a type of novelty coding for memory that maintains identity vector angle position, thereby preventing the interference of identity and novelty representations [42]. Repetition suppression magnitudes are also continuous and depend on factors such as the time since an image has been viewed, the duration of the viewing, and the number of repeated viewings [reviewed by 46]. This encoding scheme can account for behavior on a familiarity task with a decoder that maps IT neural response to behavior via a simple positively-weighted linear read-out [42].

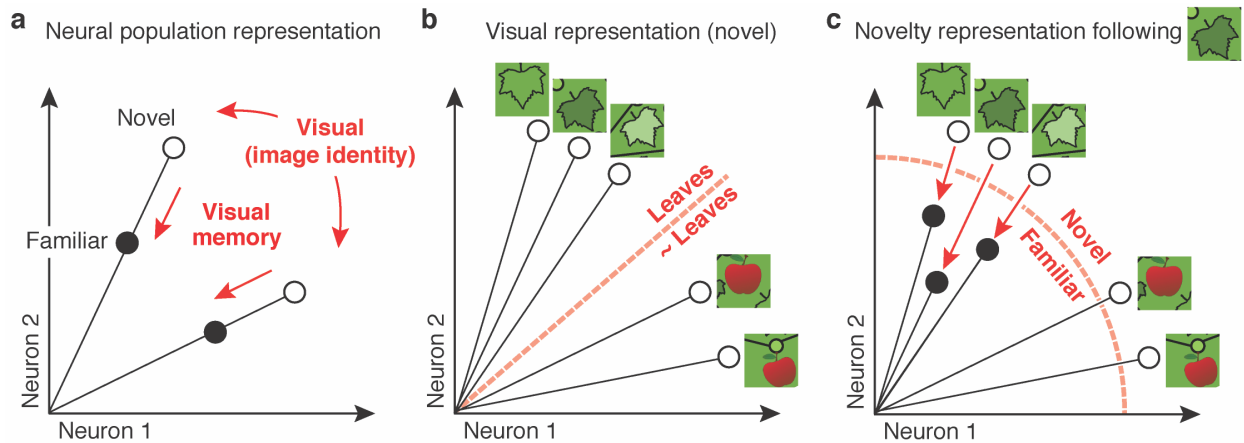

**Fig 2.** *The representation of visual novelty in IT.* **a)** Shown are the hypothetical population responses to two images, each presented as both novel and familiar, plotted as the spike count response of neuron 1 versus neuron 2. In this scenario, visual identity (e.g. image or object identity) is reflected by the population response pattern, or equivalently, the direction of each population response vector. In contrast, novelty is reflected by the population vector length, where images with longer population vectors are more novel, and novelty can be extracted with a simple, positively weighted decoder. **b)** Visual representations in IT (depicted here for novel images) are grouped by object identity and category similarity and formatted such that identity and category (e.g. leaves) can be extracted with a simple linear decoder. **c)** Repetition suppression is strongest for repeated presentations of the same image but also has partial transfer across images that are similar.

How is novelty computed by the brain? In the framework described in Figure 2, this amounts to understanding the origin of IT repetition suppression. Repetition suppression is found at all stages of visual processing from the retina to IT, and it strengthens in its magnitude as well as the duration over which it lasts across the visual cortical hierarchy [47]. Consequently, a hierarchical cascade of feed-forward, adaptation-like mechanisms clearly contribute to IT repetition suppression [46]. There are also indications that IT repetition suppression may arise from changes in synaptic weights between recurrently connected units within IT [46,48] and/or feed-back mechanisms from higher brain areas (such as perirhinal cortex) [49,50], although the latter assertion has been the focus of some debate [reviewed by 46].

**Insights from comparisons between brain-based and machine-based approaches to novelty**

Machine learning and neuroscience arrive at novelty considerations from complementary perspectives: the former solves challenging engineering problems whereas the latter discovers the principles by which the brain operates. Despite their different objectives, they have much to learn from one another. For example, insofar as the brain is a machine that can perform these tasks, engineering insight can be gained from an understanding of how it operates. Conversely, engineering considerations can provide insights into the rationale behind specific brain implementations. Here we highlight two insights that emerge from the comparison between brain-based and machine-base approaches to novelty that we find particularly striking.

*The signals for novelty are continuous and are multiplexed with identity information in the brain as well as in a subset of machine-based approaches.* As described above, the brain does not parcellate the signaling of visual identity and visual novelty into distinct modules, but rather multiplexes identity and novelty representations at the highest stages of the visual system (Fig 2a). Additionally, brain-based measures of novelty are continuous (as opposed to binary or discrete). There are interesting analogies between this distributed, continuous encoding scheme and pseudocount novelty-driven proposals, which compute continuous measures of novelty by comparing probabilities across repeated image exposures. As described above, this type of encoding scheme is highly advantageous for novelty-driven learning when the dimensionality is large and nearly everything is novel.

*The brain supports representations of visual novelty that are tolerant rather than invariant, thereby making them flexible for different tasks and environments.* To summarize one insight from above, novelty-driven RL needs to estimate view and state invariant measures of similarity between the current visual input and previously viewed images (Fig 1). In the brain, novelty information (in the form of repetition suppression) is multiplexed with visual information in IT, the same brain area implicated in the computation of object identity "invariant" to view [reviewed by 43]. While it is thus natural to conjecture that IT novelty representations are configured with the principal goal of detecting the appearance of novel objects, we suspect that this is short-sighted. Rather, as demonstrated above, the "states" that need to be explored to solve real-world learning problems (and analogously, Atari video games) are not typically defined by the objects contained therein, but rather by something more akin to scenes [see also 51]. There are several indications that IT novelty representations are configured with the broader goal of flexibly generalizing novelty information across different states. *First*, while not broadly appreciated, the responses of individual IT neurons are not themselves "invariant" but rather tolerant, meaning they remain sensitive to changes in parameters such as the position, size, and background context with which objects appear. Invariance emerges across the population as a consequence of the linearly separable format of IT representations, where object identity can be extracted with a simple linear decoding scheme (Fig 2b, [reviewed by 43]). Moreover, IT reflects explicit (i.e. linearly separable) information about scene details (such as an object's position) to a higher degree than any earlier stage of the form processing pathway [52]. *Second*, our behavioral reports of novelty contain rich information about the context and configurations that objects appear in [35], suggesting that novelty-based computations are not object-invariant, but rather incorporate considerable detail about scenes. *Third*, while novelty representations in IT are generally very selective for image identity [53], after viewing an image, repetition suppression does in fact transfer to other, similar images ([54], Fig 2c). This includes transfer across images whose similarity is conferred by their shape, as well as other variables such as an object's

spatial position. Taken together, these results suggest that graded IT novelty representations reflect information about the images that have been encountered but also support novelty generalization across images that are similar in virtue of either their view or state (Fig 1). We speculate that the representation of identity and novelty in IT (as illustrated in Fig 2a-c) supports adaptation to new environments by enabling downstream behavioral decoding to reflect the relationship of new images to those encountered in past experience.

**Conclusions**

The major successes in AI have largely come in settings with frequent, identifiable rewards, but animals live in complex environments that do not provide ubiquitous, easily identifiable rewards. Yet infants, and other developing animals, are able to reliably produce complex behaviors with only a small amount of experience. The success of novelty-based approaches in overcoming the limitations of current approaches to learning suggests that novelty computation may play a crucial role in how animals explore and learn. Signals reflecting visual novelty are found in the primate brain in higher visual areas such as IT, where they are reflected as continuous measures of novelty and multiplexed with representations of visual identity. Taken together, recent studies suggest that IT novelty representations can generalize novelty across different views and states in the manner required to drive novelty-based exploration and learning.

**Highlighted references:**

* Tang H, Houthooft R, Foote D, Stooke A, Chen OX, Duan Y, Schulman J, DeTurck F, Abbeel P: **#Exploration: A study of count-based exploration for deep reinforcement learning**. In *Conference on Neural Information Processing Systems (NeurIPS)*: 2017:2753-2762.
> This paper addressed the view and state invariance challenges by adapting count-based methods to image-based RL using hash functions that map images to a relatively small number of bins.

** Bellemare M, Srinivasan S, Ostrovski G, Schaul T, Saxton D, Munos R: **Unifying count-based exploration and intrinsic motivation**. In *Conference on Neural Information Processing Systems (NeurIPS)*: 2016:1471-1479.
> This paper introduced the notion of pseudocounts, which generalize count-based methods for novelty estimation and can be applied to RL problems with high-dimensional, continuous states such as images. The authors formally demonstrated the relationship between pseudocounts and information gain, suggesting that pseudocounts may lead to near-optimal exploration behavior.

* Savinov N, Raichuk A, Marinier R, Vincent D, Pollefeys M, Lillicrap T, Gelly S: **Episodic curiosity through reachability**. In *International Conference on Learning Representations (ICLR)*: 2019. [19]
> This paper computed an intrinsic reward signal that closely resembles a novelty computation using two interesting model components: a learned similarity measure and a memory buffer.

** Meyer T, Rust NC: **Single-exposure visual memory judgments are reflected in inferotemporal cortex**. *eLife* 2018, **7**:e32259.
> This paper demonstrated the plausibility of IT repetition suppression as a novelty signal by illustrating that a positively weighted linear read-out of IT responses could account for remembering and forgetting behavior as a function of time since an image was viewed.

* Hong H, Yamins DL, Majaj NJ, DiCarlo JJ: **Explicit information for category-orthogonal object properties increases along the ventral stream**. *Nat. Neurosci.* 2016, **19**:613-622.
> This paper demonstrated the robust and easily accessible representations of IT populations for properties beyond object identity, including object position.